# RELATING BIOPHYSICAL PROPERTIES ACROSS SCALES

**Elijah Flenner[*], Francoise Marga[*], Adrian Neagu[*†], Ioan Kosztin[*] and Gabor Forgacs[*]**

[*]*Department of Physics and Astronomy, University of Missouri, Columbia, MO 65211*
[†]*University of Medicine and Pharmacy Timisoara, 300041 Timisoara, Romania*

## Abstract

A distinguishing feature of a multicellular living system is that it operates at various scales, from the intracellular to organismal. Genes and molecules set up the conditions for the physical processes to act, in particular to shape the embryo. As development continues the changes brought about by the physical processes lead to changes in gene expression. It is this coordinated interplay between genetic and generic (i.e. physical and chemical) processes that constitutes the modern understanding of early morphogenesis. It is natural to assume that in this multi-scale process the smaller defines the larger. In case of biophysical properties, in particular, those at the subcellular level are expected to give rise to those at the tissue level and beyond. Indeed, the physical properties of tissues vary greatly from the liquid to solid. Very little is known at present on how tissue level properties are related to cell and subcellular properties. Modern measurement techniques provide quantitative results at both the intracellular and tissue level, but not on the connection between these. In the present work we outline a framework to address this connection. We specifically concentrate on the morphogenetic process of tissue fusion, by following the coalescence of two contiguous multicellular aggregates. The time evolution of this process can accurately be described by the theory of viscous liquids. We also study fusion by Monte Carlo simulations and a novel Cellular Particle Dynamics (CPD) model, which is similar to the earlier introduced Subcellular Element Model (SEM; (Newman, 2005)). Using the combination of experiments, theory and modeling we are able to relate the measured tissue level biophysical quantities to subcellular parameters. Our approach has validity beyond the particular morphogenetic process considered here and provides a general way to relate biophysical properties across scales.

# I. Introduction

In most organisms, embryonic development starts with a more or less spherical zygote. The end product of a series of morphogenetic transformations is anything but spherical. Moreover, the internal structure of the adult, composed of a multitude of organs with varying shape, interwoven by tubes and held together by the extracellular matrix, in itself is miraculously complex.

Morphogenetic shape changes require the coordinated movement of cells, which in turn requires physical forces. Movement and forces can be characterized in terms of well-defined concepts and physical parameters, such as diffusion, elasticity, viscosity, friction, velocity, etc. (Forgacs and Newman, 2005). Thus, cells also must possess physical characteristics and these, indeed, have been measured for various cell types. On the other hand cells carry out their functions according to instructions generated by their genome, a purely biochemical entity. It is the genes that set up cellular physical properties through their control of protein synthesis, the major "workers" within the cell. Changes in the number, variety and structure of proteins result in vastly differing cellular properties. An example is the actin cytoskeleton of a locomoting cell. In order to translocate, the cell needs to build a motile apparatus by the continuous polymerization/depolymerization of its filamentous actin (i.e. F actin) network. The organization of the actin cytoskeleton is thus quite different in a moving, as opposed to resting cell and this difference is manifest in distinct intracellular viscoelastic properties.

As development proceeds individual cells organize into tissues, which themselves greatly differ in physical properties: blood is liquid, bone is solid. In between these extremes lie most of the organs and tissues with typically intermediate viscoelastic properties. However, a blood cell is not the same as a liquid drop, a bone-forming cell (i.e. osteoblast) itself is not a solid and a single cardyomiocyte has totally different physical properties than the mature heart. How do intracellular and cellular physical properties determine the physical attributes of tissues, structures composed of a large number of individual cells? From the physical point of view the question may seem innocent: provided the short range interactions between cells are known, macroscopic collective tissue properties follow from the application of statistical mechanics. However, the situation is considerably more complicated when the true biological nature of cells is taken into account. Cells can produce "stuff" and thus change their interactions, just to mention one complication. Nevertheless it is intuitively obvious that ultimately the biophysical properties of tissues must derive from intracellular and cellular properties.

This article aims at elucidating how cell and subcellular physical properties may give rise to tissue level properties. We begin by developing a theoretical and computational framework to address this question. Next we apply the formalism to the fusion of tissue fragments, a morphogenetic process analogous to the coalescence of liquid drops. Specifically, we relate surface tension and viscosity, parameters characterizing fusion, to those that describe physical processes inside the cell and between cells.

## II. Theory and Computer Modeling

### A. Tissue liquidity

A careful scrutiny of early embryonic development led Steinberg to formulate the differential adhesion hypothesis (DAH) (Steinberg, 1963), which attributes morphogenetic events to differences in the cell adhesion apparatus of the different cell types. DAH implies that early morphogenesis is a self-assembly process (Whitesides and Boncheva, 2002), whereby mobile and interacting subunits spontaneously give rise to structure (Foty and Steinberg, 2005; Gonzalez-Reyes and St Johnston, 1998; Perez-Pomares and Foty, 2006; Steinberg, 1970). In light of DAH, embryonic tissues mimic the behavior of highly viscous, incompressible liquids (Steinberg and Poole, 1982). Their liquid-like behavior is indeed manifest in the rounding-up of initially irregular tissue fragments, the fusion of two or more contiguous tissue droplets into a single cellular spheroid (Gordon et al., 1972), the engulfment via spreading of one tissue type over the surface of another (Steinberg and Takeichi, 1994), and the segregation or sorting of various cell types in heterotypic cell mixtures (Foty et al., 1994; Technau and Holstein, 1992). All these phenomena can be interpreted using the theory of ordinary liquids. For example, in the absence of external forces a liquid droplet assumes a spherical shape (just as an originally irregular tissue fragment) because its constituent molecules attract each other and adopt positions that maximize their contact and minimize the overall surface area. Immiscible liquids, initially randomly intermixed, separate (similarly to the sorting of heterotypic cells) into a configuration with the more cohesive liquid being surrounded by the less cohesive one (e.g. water drop surrounded by an oil drop). It is however, important to keep in mind that true liquid molecules move due to thermal agitation, whereas cellular motion is powered by metabolic energy. Thus, the properties of embryonic tissues are analogous, but not identical to those of liquids. The physical properties that best characterize liquids are surface or interfacial tension ($\gamma$) and viscosity ($\eta$). Such quantities can effectively be attributed to embryonic tissues through measurement techniques employed in case of liquids (Forgacs et al., 1998; Foty et al., 1994; Foty et al., 1996; Gordon et al., 1972). Indeed, numerous embryonic tissues have been characterized in terms of effective tissue surface tension and the measured tensions were found to be consistent with the mutual sorting behavior of these tissues (Foty et al., 1996), as predicted by DAH. The latter provides the molecular basis of tissue surface tension by postulating a connection between this macroscopic quantity and the strength of adhesion between cells constituting the tissue. On theoretical grounds it was established that tissue surface tension has to be proportional to the number of cell surface adhesion molecules (Forgacs et al., 1998), a conclusion confirmed later experimentally (Foty and Steinberg, 2005). These findings imply a connection between tissue and cellular level quantities. The implications of DAH have been confirmed not only in vitro and in silico (Glazier and Graner, 1993a; Glazier and Graner, 1993b), but also in vivo (Godt and Tepass, 1998; Gonzalez-Reyes and St Johnston, 1998).

DAH provides useful predictions on equilibrium tissue configurations, but cannot address the question of how these configurations are arrived at in time. Since embryonic development is all about changing shapes, the more complete quantitative understanding of early morphogenetic processes necessitates a dynamical approach. It is an intriguing question whether the liquid analogy can be extended beyond equilibrium. Liquid molecules and cells both interact via short-range forces that can be characterized by their strength ($\varepsilon$) and range ($\delta$). Dynamical behavior is governed by Newton's second law, which in case of cells (due to the irrelevance of inertial forces, (Odell et al., 1981)) simplifies to frictional (i.e. Langevin) dynamics. The macroscopic properties of liquids, such as $\gamma$ and $\eta$ can be determined from the intermolecular forces in terms of $\varepsilon$ and $\delta$ (Israelachvili, 1992). Thus, if tissue liquidity remains a useful concept in dynamical developmental processes then measurable tissue level physical parameters (tissue surface tension and viscosity) could possibly be related to the strength and range of interaction between cell adhesion molecules (CAMs), such as cadherins (Takeichi, 1990). Since CAMs typically are transmembrane proteins with attachment to intracellular organelles (e.g. cadherins associate with the actin cytoskeleton (Gumbiner, 1996)), the liquid analogy, in principle, could open the possibility to establish connection between intracellular molecular entities and tissue level macroscopic physical parameters. In what follows we develop two theoretical/computational approaches to carry out this program.

**B. The Monte Carlo method**

The DAH has inspired several theoretical models of living tissues. According to DAH morphogenesis results from the movement of cells seeking positions that lead to a minimum of the total energy of adhesion (Steinberg, 1963; Steinberg, 1996). Early lattice models based on DAH yielded important insight into cellular pattern formation (Leith and Goel, 1971). Computer simulations based on these models were limited by the available computer power and the implementation of deterministic motility rules (Goel and Rogers, 1978; Rogers and Goel, 1978). In contrast, the Metropolis algorithm (Metropolis N, 1953), based on purely stochastic motility rules, turned out to be a convenient and fast method for finding conformations emerging by energy minimization. Its effectiveness in describing tissue liquidity has been shown in Monte Carlo simulations based on the Potts model, a widely used model in statistical physics (Glazier and Graner, 1993a; Graner and Glazier, 1992). In this model the tissue is represented on a lattice. Each cell is composed of several contiguous lattice sites (i.e. subcellular elements) that are labeled by an identification number (the same for each subcellular element) and a cell-type index. The average number of sites per cell is maintained around a target value. Deviations from this target value are constrained by an elastic term in the overall energy assigned to the system. Cells interact with their close neighbors. Evolution, described by the Metropolis algorithm, accounts for cell migration and shape changes resulting from the movement of subcellular elements (Glazier and Graner, 1993a). This approach was successfully applied to cell sorting and the mutual engulfment of adjacent tissue fragments. Interestingly, it was shown that cell motility could be associated to an effective temperature-like parameter (Mombach et al., 1995).

Inspired by the approach of Glazier and Graner, we have constructed a three-dimensional lattice model that enabled us to simulate tissue liquidity in systems composed of large numbers ($10^5$ to $10^6$) of interacting cells (i.e. model tissue; (Jakab et al., 2004; Neagu et al., 2005). In this model cells and similar-sized volume elements of medium or extracellular matrix (ECM) are represented as particles on sites of a cubic lattice. The type of the particle on a given site r is specified by an integer $\sigma_r \in \{1, 2, ..., T\}$. The contact interaction energy $J(\sigma_r, \sigma_{r'})$ between two particles, located at neighboring sites $r$ and $r'$, of type $\sigma_r = i$ and $\sigma_{r'} = j$, respectively, is given by $J(\sigma_r, \sigma_{r'}) = -\varepsilon_{ij}$, where $\varepsilon_{ij}$ is the mechanical work needed to break the bond between them. For $i \neq j$ ($i = j$) the energy $\varepsilon_{ij}$ is referred to as the work of cohesion (adhesion). The total interaction energy of the model tissue is written as

$$E = \sum_{\langle r,r'\rangle} J(\sigma_r, \sigma_{r'}), \tag{1}$$

where $\langle r, r' \rangle$ indicates that the summation involves only close neighbors. Specifically, we consider interactions between nearest, next-nearest and second-nearest neighbors. Thus, the total number of neighbors interacting with a given particle is $n = 26$. For simplicity we assume that neighbors of the same type interact with the same strength, irrespective whether they are nearest, next-nearest and second-nearest neighbors. Thus, in the case of a homocellular tissue fragment (particles of type 2) in cell culture medium (particles of type 1), $J(\sigma_r, \sigma_{r'})$ may take either of the values $J(1,1) = -\varepsilon_{11}$, $J(2,2) = -\varepsilon_{22}$, and $J(1,2) = J(2,1) = -\varepsilon_{12}$. Note that in this case $\varepsilon_{11} \ll \varepsilon_{22}$ and $\varepsilon_{12} \ll \varepsilon_{22}$.

By separating interfacial terms in the sum, the adhesive energy of a system made of $T$ types of particles, up to an irrelevant additive constant, becomes (Neagu et al., 2006)

$$E = \sum_{\substack{i,j=1 \\ i<j}}^{T} \gamma_{ij} N_{ij}, \tag{2}$$

provided that particles do not change their type. Here $N_{ij}$ denotes the total number of bonds between particles of type $i$ and type $j$ ($i \neq j$). It is noteworthy that the total interaction energy depends only on the number of heterotypic bonds, $N_{ij}$ and the interfacial tension parameters (Jakab et al., 2004)

$$\gamma_{ij} = \tfrac{1}{2}(\varepsilon_{ii} + \varepsilon_{jj}) - \varepsilon_{ij}. \tag{3}$$

The model can readily be extended to include cell differentiation, proliferation or death, as well as the remodeling of the ECM by cells. In such situations the interaction energy in Eq.1 depends explicitly on the "works of cohesion"

$$E = \sum_{\substack{i,j=1 \\ i<j}}^{T} \gamma_{ij} N_{ij} - \frac{1}{2} n \sum_{i=1}^{T} N_i \varepsilon_{ii}. \tag{4}$$

Now $N_i$ (number of type $i$ particles in the system) may vary during the simulation.

We have applied the Metropolis Monte Carlo method (Amar, 2006; Metropolis N, 1953) to the above model to investigate the evolution of a cellular system that displays random motility and is consistent with DAH. Specifically, the employed computational algorithm consisted of the following steps.

1. The initial state of the system is constructed by assigning to each lattice site a cell type index that specifies occupancy.
2. Particles that are in contact with other type of particles, and thus represent interfacial cells, are identified.

3. Each interfacial cell is given the chance to move by exchanging its position with a randomly selected neighbor of a different type. The corresponding energy change $\Delta E$ is computed after each move and the new configuration is accepted with a probability $\min(1, \exp(-\Delta E/E_T))$. Thus an energy-lowering movement is accepted with probability one. Changes in position that lead to an increase in energy are also allowed, albeit with smaller probability given by the Boltzmann factor.

4. The configuration, the interaction energy, and the values of interfacial areas are written into output files.

5. Interfacial cells are identified in the new configuration and the process is repeated from step 3 until the desired number of steps is completed.

The energy $E_T$ in the expression for the acceptance probability is an effective measure of cell motility. It is the biological analogue of $k_B T$, the thermal fluctuation energy ($k_B$ - Boltzmann constant, $T$ - absolute temperature). Being related to cytoskeleton-driven cell membrane ruffling (Mombach et al., 1995), $E_T$ is referred to as the biological fluctuation energy. It has experimentally been assessed for certain embryonic cell types (Beysens et al., 2000). The interaction energies in Eqs. 1-3 are expressed in units of $E_T$.

A Monte Carlo step (MCS) is defined as the set of operations during which each interfacial cell has been given the chance to experience a change. The simulations were implemented with fixed boundary conditions by confining cell movement to the volume of the simulated region. The concern that finite-size effects may interfere with the energetically-driven rearrangement of cells is justified only when a significant fraction of the cells migrates away from the model tissue construct and reaches the system boundary. This is usually not the case in our simulations.

## C. Cellular Particle Dynamics method

A more realistic approach to simulate the motion and self-assembly of large cell aggregates is to model the individual cells as a set of interacting *cellular particles* (CPs), and follow their 3D spatial trajectories in time by integrating numerically the corresponding equations of motion in the spirit of Newman's Subcellular Element Model (SEM (Newman, 2005)). CPs result from a coarse-graining description of the cells. The number and type of the CPs that constitute a cell, are determined by the chosen level of coarse-graining, makes the model flexible. The motion of a cell, as well as dynamical changes in its shape, are determined by the collective motion of its constituent CPs.

At a given time the location of a CP is determined by its position vector, $r_{i,n}(t)$, where the index $i = 1,...,N_{CP}$ labels the CPs within the cell. Here we consider a system of $N$ cells labeled by the index $n$ ($n = 1,...,N$). The precise form of the interaction between CPs is dictated by the detailed biology of the cell, whose complexity makes the determination of this interaction potential practically impossible. However, it is quite reasonable to assume that CPs interact through local biomechanical forces which (similarly to liquid molecules) can conveniently be modeled by the Lennard-Jones (LJ) potential energy

$$V_{LJ}(r; \varepsilon, \sigma) = 4\varepsilon \left[ \left(\frac{\sigma}{r}\right)^{12} - \left(\frac{\sigma}{r}\right)^{6} \right], \tag{5}$$

where $r$ is the distance between the CPs, $\varepsilon$ is the energy required to separate the CPs and $\sigma$ gives the size (diameter) of a CP. The *inter-cellular* interaction potential between two CPs that belong to two different cells is $V_2(r) = V_{LJ}(r;\varepsilon_2,\sigma_2)$. If the two CPs belong to the same cell, the corresponding *intra-cellular* potential has the form $V_1(r) = V_{LJ}(r;\varepsilon_1,\sigma_1) + V_c(r;k,\alpha)$. Here $V_c$ represents a *confining* potential energy of the form

$$V_c(r;k,\alpha) = \begin{cases} (k/2)(r-\alpha)^2, & \text{for } r > \alpha \\ 0 & \text{for } r < \alpha \end{cases} \quad (6)$$

which assures that CPs for a given cell remain inside the cell (Flenner et al.). Thus $V_c$ preserves the integrity of the cell. Indeed, whenever the separation between two CPs within a given cell exceeds a predefined distance $\alpha$ an elastic force $F_c = -\partial V_c/\partial r = -k(r-\alpha)$ is turned on to prevent the two CPs to move further apart. Thus by tuning the values of $k$ and $\alpha$ in the confining potential one can control, respectively, the stiffness and the size of a cell. While the LJ parameters corresponding to the inter- and intra-cellular interactions may have different values, here, for simplicity, we shall assume that these are equal, i.e., $\varepsilon_1 = \varepsilon_2 = \varepsilon$ and $\sigma_1 = \sigma_2 = \sigma$.     In addition, each
CP interacts with its highly viscous and stochastic environment (i.e. cytosol). These interactions can be described by a friction force $\boldsymbol{F}_f = -\mu \boldsymbol{v}$ ($\mu$ is the friction coefficient and $\boldsymbol{v} = \dot{\boldsymbol{r}}$ is the velocity of the CP), and a stochastic force $\boldsymbol{\zeta}(t)$, modeled as a Gaussian white noise with zero mean and variance $\langle \zeta_i(t)\zeta_j(0)\rangle = 2D\mu^2\delta(t)\delta_{ij}$ ($D$ is the self-diffusion coefficient of a CP). Although living cellular systems are not in thermodynamic equilibrium, we assume that the "Einstein relation" $D\mu = E_T$ still holds, where $E_T$ is the biological fluctuation energy introduced earlier. Finally, assuming that the motion of the CPs in the highly viscous medium is overdamped (i.e., inertia can be neglected), the corresponding Langevin equation of motion for the $i$-th CP in the $n$-th cell can be written as

$$\mu\dot{\boldsymbol{r}}_{i,n} = \boldsymbol{\zeta}_{i,n}(t) - \underbrace{\sum_{j\neq i}\partial V_1(|\boldsymbol{r}_{i,n}-\boldsymbol{r}_{j,n}|)/\partial\boldsymbol{r}_{i,n}}_{\boldsymbol{F}_{1i,n}} - \underbrace{\sum_{j,m\neq n}\partial V_2(|\boldsymbol{r}_{i,n}-\boldsymbol{r}_{j,m}|)/\partial\boldsymbol{r}_{i,n}}_{\boldsymbol{F}_{2i,n}}, \quad i=1,...,N_{cp}; n,m=1,...,N, \quad (7)$$

where $\boldsymbol{F}_{1i,n}$ ($\boldsymbol{F}_{2i,n}$) represents the net intra- (inter-) cellular interaction force exerted by the other CPs.

Similarly to SEM (Newman, 2005), the CPD approach consists in numerically integrating the above equations in order to determine the individual trajectories of all CPs. This can be achieved most efficiently by properly implementing the intra- and inter-cellular interaction forces and a Langevin dynamics integrator in one of the freely available massively parallel molecular dynamics (MD) packages (Flenner et al.). We have successfully implemented CPD into two such MD programs, NAMD (Phillips et al., 2005) and LAMMPS (Plimpton, 1995). We used both implementations to simulate the fusion process of two identical spherical cell aggregates as described in the Results section. The obtained results were insensitive to which MD package was used.

## D. Tissue fusion

Tissue fusion is a ubiquitous morphogenetic process (Perez-Pomares and Foty, 2006). A specific example in early development is the fusion of embryonic cushions, a process leading to the formation of the 4-chambered heart, i.e., septation (Wessels and Sedmera, 2003). We have investigated tissue fusion experimentally using roundup fragments of several cell types. When two embryonic tissue spheroids (in tissue culture medium) were arranged contiguously by gravitational forcing in hanging drop configuration (on a lid of an inverted Petri dish) they fused into a single aggregate. Figure 1 shows snapshots of fusion in the case of aggregates of smooth muscle cells. (For the preparation of such aggregates, see for example (Foty et al., 1996; Hegedus et al., 2006). To quantify the fusion process, we followed the time evolution of the interfacial area of contact between the fusing tissue drops (Fig. 1). The process was then simulated using both the Monte Carlo and the CPD methods the latter with either one (CPD-1) or ten (CPD-10) CPs per cell (Fig.1).

As will be shown, the fusion of two cellular aggregates is similar to the coalescence of two viscous liquid drops. The process is driven by surface tension $\gamma$ and is resisted by viscosity $\eta$. In principle, the precise theoretical description of shape evolution during fusion can be given using the laws of hydrodynamics. Under some reasonable approximations applicable to highly viscous incompressible liquids (Frenkel, 1945), we have found an exact solution to the problem, which is in excellent agreement with both our experimental and computer simulation results (Flenner et al., 2007). In our analytical model, the fusing drops are spherical caps whose radii increase as their centers move closer together, eventually forming a single spherical drop with radius $R_f = 2^{1/3} R_0$, determined from volume conservation. Here $R_0$ is the initial radius of the two identical spherical drops. Accordingly, the interfacial contact between the fusing drops is a circle of radius $r_0$, which increases from 0 to $R_f$. While the mathematical expression of the time dependence of $(r_0/R_0)^2$ (i.e., the area of interfacial contact $\pi r_0^2$ expressed in units of $\pi R_0^2$), is rather complicated, the solution is almost indistinguishable from the simple formula

$$(r_0/R_0)^2 = 2^{2/3}(1 - e^{-t/1.35\tau_0}), \qquad \tau_0 = \eta R_0/\gamma. \tag{8}$$

Because $(r_0/R_0)^2$ approaches exponentially its limiting value $(R_f/R_0)^2 = 2^{2/3}$, with a time constant $\tau = 1.35\tau_0$, the definition of the total fusion or rounding time $t_R$ is somewhat arbitrary. We define $t_R$ as the instant of time when $r_0$ reaches 96% of the final radius $R_f$, i.e., $r_0(t_R) = 0.96 R_f$. At this time, for all practical purposes, the fused aggregates are completely rounded up (see also Fig.1). From Eq.8 it follows that $t_R$ is related to the characteristic time $\tau_0$ through the formula $t_R = 3.45\tau_0 = 3.45\eta R_0/\gamma$. Furthermore, as discussed later, Eq.8 allows relating tissue level quantities to molecular, cell and subcellular quantities.

## III. Results

### A. Monte Carlo simulations

The Monte Carlo method was used to study a number of quantities characterizing the fusion of two cellular aggregates, such as the surface energy and the degree of mixing of cells. Snapshots of simulated structural changes in the model system are shown in Fig. 1.

The Monte Carlo simulations presented in this article were performed with the "works of adhesion" $\varepsilon_{cc} = 1$ and $\varepsilon_{mm} = 0$, describing respectively cell-cell interactions and medium-medium interactions and $\varepsilon_{cm} = 0$, the associated cell-medium interaction (all interactions are expressed in units of the cellular fluctuation energy, $E_T$). This choice of parameters corresponds to a cellular aggregate placed in cell culture medium (with which its interaction is negligible) and results in a cell-aggregate–medium interfacial tension $\gamma_{cm} = 0.5$ (see Eq. 3).

Figure 2 illustrates how the interfacial energy (and thus the surface area), and thereby the surface area, of the model tissue decreases during fusion. The lowering of the interfacial energy continues until it levels off, showing that the fusion and complete rounding of the initial spheroids occurs in about $1.2 \times 10^6$ MCS. As shown in the inset of Fig. 2, fluctuation-driven deviations from the spherical shape of the initial aggregates cause a slight increase in surface area right after the start. The interfacial energy begins to decrease as soon as the two spheroids make contact and start to coalesce. Figure 2 demonstrates the efficiency of the Metropolis algorithm in selecting energetically favorable rearrangements of motile, interacting particles (thus mimicking liquidity despite the fact that no real time evolution is being considered). An important feature of our implementation of the Metropolis algorithm is that it only includes conformational changes that result from the movement of cells. This is in contrast with Monte Carlo simulations based on completely random changes in particle arrangement.

According to the first two columns in Fig. 1, Metropolis Monte Carlo (MC) simulations qualitatively reproduce the experimental sequence of intermediate stages observed during fusion. We also investigated if this similarity pertains as well to the movement of individual cells within the fusing aggregates. To this end, we studied how cells originating from one of the spheroids diffuse into the other one. The simulations started with two contiguous aggregates made of model cells of the same type, but labeled and colored differently for bookkeeping (Fig. 3). The system was "sectioned" normally to the longitudinal axis of symmetry of the initial configuration. The position of each transversal plane was specified by the coordinate $x$ of its intersection with the longitudinal axis. The fractions $f_L$ and $f_R$ of cells initially located in the left and right aggregates were respectively defined by

$$f_\beta(x) = \frac{n_\beta(x)}{n_L(x) + n_R(x)}, \qquad \beta = L, R, \qquad (9)$$

where $n_L(x)$ ($n_R(x)$) is the number of cells originating from the left (right) aggregate and located in the cross-section of coordinate $x$. In our lattice formulation, the coordinate $x$ only takes on the discrete values $x_s$, with $s = 1, 2, \ldots, S$, with $S$ being is the total number of those cross-sections that contain model cells.

As a global measure of particle mixing in a given configuration, we define the degree of mixing $d_m$ by the relation

$$d_m = \frac{1}{N} \int f_L(x) f_R(x) dx = \frac{1}{N} \sum_{s=1}^{S} f_L(x_s) f_R(x_s). \tag{10}$$

The normalization factor $N$ is chosen such that $d_m = 1$ for complete mixing, when $f_L(x) = f_R(x) = 0.5$ in each cross-section that contains cells. In the discrete formulation this condition yields $N = S/4$.

The curves in Fig. 3 represent the variation of $f_L$ along the symmetry axis of the system at different instants of time. In the initial state, $f_L(x) = 1$ for $x < 0$ and $f_L(x) = 0$ for $x > 0$. As the fusion evolves, cells mix and the graph of $f_L(x)$ flattens, approaching the value 0.5 associated with a degree of mixing $d_m = 1$. Figure 3 shows that within $3 \times 10^3$ MCS the cells starting from different initial aggregates mix completely, though the fusion is still in an incipient stage (see the snapshot of the corresponding state in the lower left corner of Fig. 3). The plot of the degree of mixing versus the number of elapsed MCS (Fig. 4) also supports the conclusion that over 90% mixing occurs within about 0.1% of the number of MCS needed for complete rounding. This finding is at variance with experiments (i.e. confocal microscopic analysis of the series in Fig. 1) showing negligible cell mixing until the aggregates coalesce forming a cylindrical shape with hemispherical caps.

During simulations of fusion we monitored the radius of the circular contact area of the two spherical aggregates. Using the method illustrated in Fig. 5, we calculated the surface density of cells in this plane for intermediate states of the simulation as a function of the distance $r$ from the center of symmetry of the initial system. It was found that this density is well described by the function

$$\rho(r) = \frac{1}{2}\left[1 - \tanh\left(\frac{2(r-r_0)}{b}\right)\right], \tag{11}$$

which is also used in analyses of simulations of spherical liquid drops (Thompson et al., 1984). We identify the parameter $r_0$ in the fit as the radius of the contact area. The second fitting parameter, $b$, is related to the steepness of the density profile in the vicinity of the surface.

In Fig. 6 the evolution of $(r_0/R_0)^2$ is plotted as a function of the elapsed MCS. The data points in Fig. 6 represent the mean values of $(r_0/R_0)^2$ obtained by averaging the results of 30 simulations started with different seeds of the random number generator. After about $1.2 \times 10^6$ MCS complete rounding takes place and $(r_0/R_0)^2$ reaches its limiting value $(R_f/R_0)^2 = 2^{2/3} \approx 1.59$, in agreement with volume conservation. Furthermore, according to Fig. 2, at this stage of the simulation the surface area and interfacial energy approach their minimum.

We conclude this section by noting that Monte Carlo simulations based on a lattice representation of a living tissue qualitatively describe the liquid-like behavior of embryonic tissues. However, discrepancy between the experimental and simulated mixing of cells during aggregate fusion shows that the Metropolis algorithm is inappropriate for describing individual cell motility in a three-dimensional tissue. Another shortcoming of the Metropolis Monte Carlo method is that apriori there is no relationship between real time and the number of performed MCS.

## B. Cellular Particle Dynamics (CPD) simulations

Here we present results of two CPD simulations for the fusion of two cellular aggregates. In the first simulation, referred to as CPD-1, each cell was formed by a single CP, whereas in the second simulation, referred to as CPD-10, each cell contained ten CPs. While both models lead to liquid-like behavior, each model has advantages and disadvantages. We first describe the simulations in detail. Then, based on the analysis of the obtained CPD trajectories, we calculate the same quantities as in the MC simulations, i.e., the degree of mixing of the cells and the time evolution of the contact area between the drops. For convenience, we employ as length, energy and time units respectively $\sigma$, $E_T$ and $\sigma^2 \mu / E_T$. In terms of these units the CPD equations of motion (Eq.7) are equivalent to setting $\sigma = E_T = \mu = 1$ in them. Thus, we only need to specify the dimensionless values for $\varepsilon$, $k$ and $\alpha$ (the latter two enter only in CPD-10).

In CPD-1 the LJ energy parameter was set to $\varepsilon = 2.5$. This relatively large energy value was motivated by the need to maintain the integrity of the cell aggregates against the escape of cells located at their surface. To make the LJ interaction between cells truly short ranged it was truncated at two length units. The starting configuration of the spheroidal cell aggregates was generated as follows. First, from a compact cubic lattice of cells a spherical aggregate (centered about the origin of the lattice) of radius $R_0 = 11$ was created, by removing all the cells situated outside this sphere. Next, two copies of this spherical aggregate were placed such that the distance between their centers was $2R_0$. Initially, each of the aggregates contained 2,108 cells, and the two aggregates had a small but finite contact region, which facilitated the fusion process right at the beginning of the CPD simulation. We used periodic boundary conditions with a simulation box of side $300\sigma$. During the simulation only a few cells crossed the simulation box. The Langevin CPD equations were integrated using the Euler algorithm with a time step $\Delta t = 10^{-5}$.

In the CPD-10 simulation the LJ energy parameter was set to $\varepsilon = 1$, which is smaller than in the CPD-1 case. The integrity of the individual cells was enforced by the confining potential with $\alpha = 2.5$ and $k = 5$. Similarly to CPD-1, a spherical aggregate of 200 cells was constructed on a lattice and then equilibrated through a CPD run until a steady state, with small energy fluctuations, was reached. The initial configuration of the system was built from two contiguous equilibrated aggregates.

A sequence of snapshots of the fusion process for both CPD simulations is shown in Fig.1. In agreement with our theoretical prediction of the time evolution of the fusion process, the shapes of the fusing aggregates are similar for both experiment and simulations. The rounding time was found by examining the time evolution of $r_0$ (defined in Eq.11). Note that in CPD-1 $r_0$ had a finite value at the start of the simulation. The snapshots show that both CPD models reproduce the experimentally observed fusion process. However, a notable difference between the two simulations is that while in CPD-1 by the end of the fusion process approximately 10% of the cells escaped from the two aggregates, in case of CPD-10 none of the cells detached. Since maintaining the integrity of the system in case of CPD-1 is problematic, the general conclusion is that in CPD simulations cells must be built from at least a few CPs. The actual choice of the number of CPs depends on the problem at hand.

Another important difference between the two CPD simulations concerns the degree of mixing of the (otherwise identical) cells from different aggregates, which is noticeably larger in case of CPD-1. To examine the degree of mixing we have used the continuum version of Eq. 9,

$$f_\beta(x,t) = \frac{\sum_{i=1}^{N_\beta} \delta(x - x_i(t))}{\sum_{i=1}^{N} \delta(x - x_i(t))}, \qquad \beta = L, R, \tag{12}$$

where $x_i(t)$ is the position of particle $i$ at time $t$ along the $x$-axis (as defined earlier). In the numerator for $\beta = L$ ($R$) the sum is over particles that originate from the left (right) aggregate, whereas in the denominator the sum is over all particles. Here $\delta(x)$ stands for the Dirac delta function (i.e., $\delta(x) = 0$ except $x = 0$, and $\int_{-\infty}^{\infty} \delta(x) dx = 1$). In Fig. 7 we plot $f_L(x,t)$ for the two CPD simulations. The different curves correspond to three different snapshots, $t = 0$, $0.5 t_R$ and $t_R$. Recall that in CPD-1 the fusion process has already started at $t = 0$, meaning that the degree of mixing in this case is underestimated. However, contrary to the Monte Carlo simulations, complete mixing does not occur by the end of the fusion process even in the case of CPD-1.

The degree of mixing $d_m(t)$, given by Eq. 10, was calculated for both simulations. As the results in Fig. 8 show, the rate of mixing appears to be constant throughout both simulations. The rate of mixing, however, is noticeably smaller for CPD-10 than CPD-1.

From the CPD trajectories, the time evolution of the relative contact area between the two fusing aggregates $(r_0/R_0)^2$ was also determined, as shown in Fig. 9. The algorithm for identifying the contact area and its radius $r_0$ was similar to the one employed in the case of MC simulations. The (dimensionless) rounding time $t_R = N_{t_R} \Delta t$ was defined as the smallest time $t$ when $r_0(t) = R_f$, the average final (stationary) radius of the fused system.

## C. Comparison of Experiments, Theory and Computer Simulations

Figure 1 shows the comparison of shape evolution of the fusing aggregates as obtained in the experiments, the various simulations and by the theoretical analysis of the process. In order to quantitatively compare the various methods described we use the time dependence of $(r_0/R_0)^2$. Experimental data on this quantity was obtained by following the fusion of two aggregates composed of embryonic smooth muscle cells. Results of the comparison are shown in Fig. 10.

Equation 8 was used to fit the data. This provided the ratio $\eta/\gamma$ for the tissue, but, more importantly, also allowed connecting the macroscopic, tissue level quantities ($\gamma, \eta$) to the molecular, cell and subcellular quantities that characterize the CPD model (i.e., $\varepsilon, \sigma$ - parameters in the LJ potential, $k, \alpha$ - parameters in the confining potential, and $\mu, E_T$ - cytosolic, "environmental" parameters). A practical way to implement this connection is as follows. As in any particle dynamics simulation we need to choose proper dimensionless units. By measuring length, energy and time in units of $\sigma$, $E_T$ and $\sigma^2/D = \sigma^2 \mu/E_T$, respectively, amounts to formally setting $\sigma = E_T = \mu = 1$ in the CPD equations of motion. Furthermore, it is natural to choose the range of the confining potential to be the size of the cell, i.e., $\alpha \sim d \sim N_{CP}^{1/3} \sigma$, where $N_{CP}$ is the number of CPs in a cell. In our simulation $N_{CP} = 10$ and, in dimensionless units, we set $\alpha = 2.5$. Thus, we only need to specify the values of $\varepsilon$ and $k$. Although, in principle, these can be considered as ad-justable parameters in the CPD simulation, we set $\varepsilon = 1$ and $k = 5$. Finally, choosing a suitable integration time step $\Delta t$ (typically $\sim 10^{-4} - 10^{-5}$), by starting from the initial configuration of two contiguous spherical cellular aggregates (connected through a point like region), we integrated the CPD equations of motion for $N_t$ time steps using (the modified) LAMMPS code. By visualizing the obtained CPD trajectory (e.g., by using the program VMD (Humphrey et al., 1996)) we followed the time evolution of the fusion process (showed in Fig.1). From the CPD trajectory, one can easily determine the rounding time $t_R = N_{t_R} \Delta t$ (in dimensionless units), where $N_{t_R}$ is the corresponding number of simulation steps. In case of CPD-10, the radius of the final fused spheroid $R_f$ was found to be equal (within the errors due to the fluctuations of the system) to the theoretical value $2^{1/3} \approx 1.26$ dictated by volume conservation. However, in case of CPD-1, where a noticeable fraction of the CPs escaped the aggregates during their fusion, $R_f$ was only ~1, a value consistent with the number of lost CPs. As shown in Fig.10, the contact area vs time in case of CPD-10 compares rather well both with the experimental data and the corresponding theoretical curve. The difference in the results, as well as the irregular shape of the CPD-10 curve, are due to the small system size, i.e., only 200 cells per aggregate. For a much larger system containing 2,000 cell per aggregate the corresponding CPD-10 curve (data not shown) matches very well the theoretical result. We emphasize that the $(r_0/R_0)^2$ vs $t/t_R$ curve is universal (i.e., free of any fitting parameter) as long as the fusing drops are incompressible and very viscous. Thus, once the fusion time $t_R$ is determined, the data points $(r_0/R_0)^2$ obtained from experiment or simulations should lie on the theoretical curve given by Eq.8.

The CPD trajectory provides $(r_0/R_0)^2$, as a function of $t/t_R = N_t/N_{t_R}$, whose graphical representation can readily be compared with both experimental and theoretical results (see Fig. 10). As already mentioned, in terms of $\tau_0$ the rounding time $t_R \approx 3.45 \tau_0 = 3.45 \eta R_0/\gamma$. On the other hand, in physical units, the CPD rounding time is $t_R = N_{t_R} \Delta t \sigma^2 \mu / E_T$. Equating these two expressions allows to determine $\mu$, a cellular level quantity (that characterizes the viscous properties of the cell's interior, i.e. cytosol) in terms of tissue level properties that are accessible by biophysical measurements. As already mentioned, $\sigma$ is determined by the known size of the cell and the number of CPs in the simulation (in our case 1 or 10).

Since the surface tension $\gamma$ is an equilibrium quantity, one expects that it is related only to the LJ parameters in the CPD model. Indeed, by calculating the energy per unit area of the CPs in the surface layer of an equilibrated spherical aggregate, one finds that $\gamma = c \varepsilon / \sigma^2$ (where $c$ is a coefficient of order unity that is only slightly model dependent), leading to yet another relationship between measurable tissue parameters and molecular quantities. Furthermore, this result implies that for two CPD models (having the same number of CPs) describing two distinct tissues,

$$\frac{\varepsilon_2}{\varepsilon_1} = \frac{\gamma_2}{\gamma_1} \left(\frac{\sigma_1}{\sigma_2}\right)^2 = \frac{\gamma_2}{\gamma_1} \left(\frac{d_1}{d_2}\right)^2. \tag{12}$$

This equation relates the LJ energy parameters of one cell type to those of another. The above approach to calculate $\gamma$ can also be used in the MC simulations. Such a calculation makes it possible to relate the model parameters in MC to those in CPD and thus to experimentally measurable equilibrium quantities.

Finally, we discuss the possibility of comparing the MC and CPD simulations results on the time evolution of the contact area. In general, there is no meaningful association between MCS and time. However, in the case of fusion of two aggregates, as shown in Fig.1, during the course of the MC simulation the system seems to go through similar intermediate configurations between the initial and final states as in the CPD simulations. This allows establishing a non-linear relationship between the MCS and time in CPD based on the shape similarity. The result is shown in the inset to Fig.10. With this correspondence it is possible to represent on the same plot in Fig.10 the "pseudo-time" dependence of the contact area as obtained from MC simulations.

## IV. Conclusions

Biological systems are extremely complex, primarily because they extend over vastly differing scales, from the molecular to the organismal. How the genetic level information stored in the sequence of nucleotides with the size of nanometers is processed and eventually implemented at the size of meters is an unresolved puzzle. It is clear that problems of such magnitude will not be possible to solve with the traditional methods of the individual scientific disciplines such as physics, biology and chemistry. Furthermore, the vast number of degrees of freedom characterizing living systems necessitates the use of powerful computational methods. Contrary to non-living physical systems, where reduction in the degrees of freedom is often possible (depending on the specific question asked), in living systems almost everything matters. Simplifying the system through the elimination of degrees of freedom may result in "throwing out the baby with the bathwater".

In the present work, combining experimental tools with theory and computer simulations we outlined the initial steps of a program to tackle the problem of multiple scales in living systems. On one hand we measured tissue level biophysical properties, such as surface tension and viscosity (in fact here we only concentrated on the ratio of these quantities). Such macroscopic material properties of living tissues must originate from cell and subcellular processes. No matter how evident this statement might be, no systematic method exists today to address the question of how cell and subcellular processes and properties (in particular mechanical ones, studied here) give rise to those at the level of tissues. To make progress, we considered a specific early developmental process, tissue fusion. Relying on the postulates of Steinberg's Differential Adhesion Hypothesis on the liquid-like nature of embryonic tissues (amply supported by in vitro and in vivo experiments), we modeled tissue fusion both theoretically and through computer simulations. The theoretical approach employed the basic laws of hydrodynamics. Computer simulations were carried out using both Monte Carlo and Cellular Particle Dynamics methods. The latter approach is similar to SEM (Newman, 2005).

The essence of our approach is that biological complexity at the molecular scale, where accurate quantitative measurements are difficult and scarce, is treated with specifically designed computational methods that connect cellular level quantities to tissue level quantities. Specifically, in our CPD model we used the Lennard-Jones potential that accurately describes intermolecular interactions in liquids to represent intra and inter-cellular interactions. Theoretical models, like the one employed here, have no problem bridging the gap between molecular and macroscopic scales. The laws of hydrodynamics provide a coarse-grained description of liquids (in terms of densities and associated velocities), thus apply to the mesoscopic/macroscopic scale.

However, they have their solid molecular foundation. We employed hydrodynamics to solve the problem of fusion of two viscous liquids and provide analytical expressions as a convenient way to compare the results of CPD and MC simulations with experimental results generated at the tissue level. The analytical expressions allowed to directly relate quantitative parameters at the cell and subcellular (i.e. molecular) level with those characterizing multicellular tissue. We believe our approach has utility beyond the specific example of tissue fusion considered here and provides a framework for further investigation on the connection of biological processes across scales.

**Acknowledgement**

This work was supported by the National Science Foundation under Grant FIBR-0526854. We gratefully acknowledge the computational resources provided by the University of Missouri Bioinformatics Consortium.

# Figures

| $t/t_R$ | EXP | MC | CPD - 1 | CPD - 10 | THEORY |
|---|---|---|---|---|---|
| 0.02 | | | | | |
| 0.09 | | | | | |
| 0.13 | | | | | |
| 0.21 | | | | | |
| 0.32 | | | | | |
| 0.45 | | | | | |
| 0.54 | | | | | |
| 0.77 | | | | | |
| 1.00 | | | | | |

**Figure 1** Snapshots of the fusion of spherical cell aggregates obtained from experiment, MC and CPD simulations (CPD-1 and CPD-10), and from theoretical modeling. The snapshots in the horizontal rows were taken at well defined instants of time (expressed in term of the rounding, or total fusion time $t_R$) listed in the first column, except the MC results which were taken after 0, 3, 12, 50, 150, 300, 500, 800 and 1200 thousands of MCS.

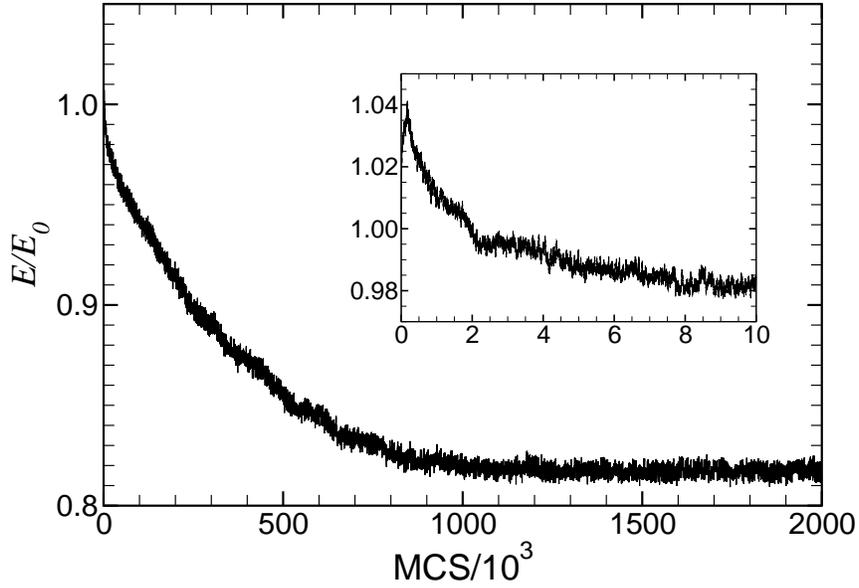

**Figure 2** Simulated evolution of the surface energy of the model tissue, expressed in units of its initial value. Here MCS stands for the number of elapsed Monte Carlo steps. The inset depicts the same dependency for the first $10^4$ MCS. The tissue-medium interfacial tension parameter was $\gamma_{cm} = 0.5$.

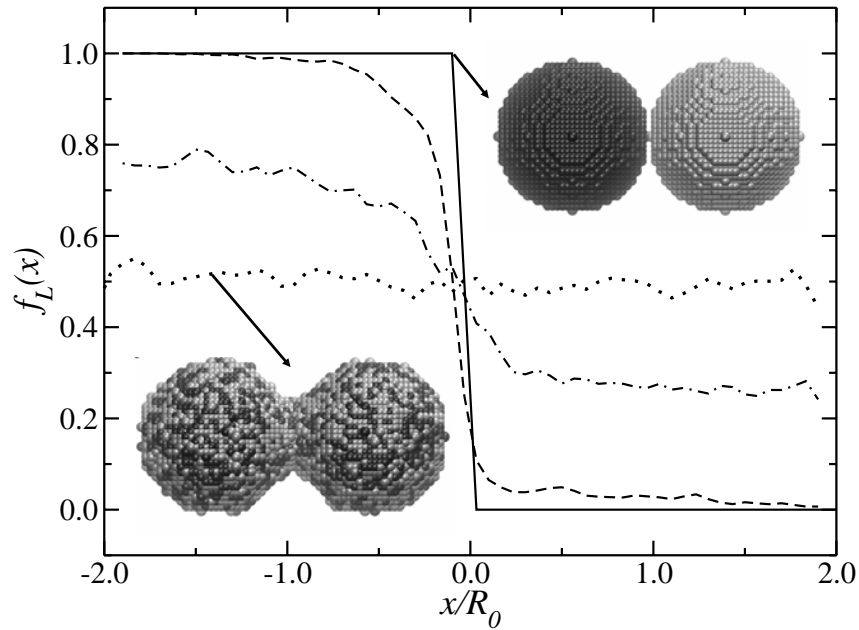

**Figure 3** The mixing of cells during aggregate fusion simulated by the Metropolis Monte Carlo algorithm. The starting system is shown in the upper right corner: two aggregates made of the same type of cells, colored differently in order to keep track of mixing. The curves plot the fraction vs. position for cells originating from the left aggregate. Cells are counted in cross-sections taken normally to the longitudinal axis of symmetry of the system ($Ox$). The coordinate of the intersection of the axis with the plane of cross-section is expressed in units of the initial aggregate radius $R_0$; the origin corresponds to the center of symmetry of the initial configuration. Each curve refers to a different stage of the simulation as follows: initial state – step function, and intermediate states at 300, 1000, and 3000 MCS – curves gradually converging to the horizontal line corresponding to complete mixing (fraction of cells from either aggregate in each cross section is 0.5). The model tissue conformation obtained in 3000 MCS is depicted in the lower left corner. The interfacial tension parameter was $\gamma_{cm} = 0.5$.

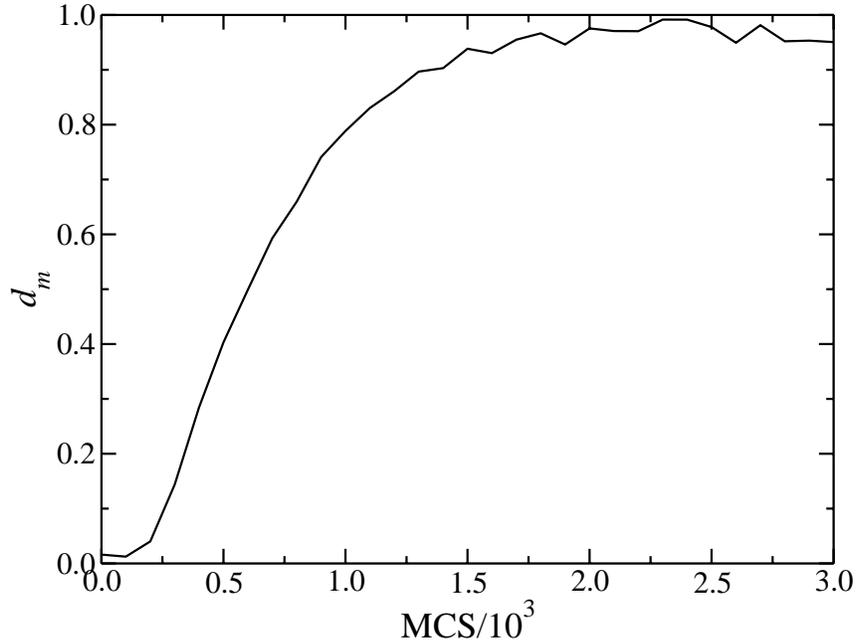

**Figure 4** Degree of mixing (Eq. 6) vs. elapsed MCS. Complete rounding occurs in about $1.2 \times 10^6$ MCS for interfacial tension parameter $\gamma_{cm} = 0.5$.

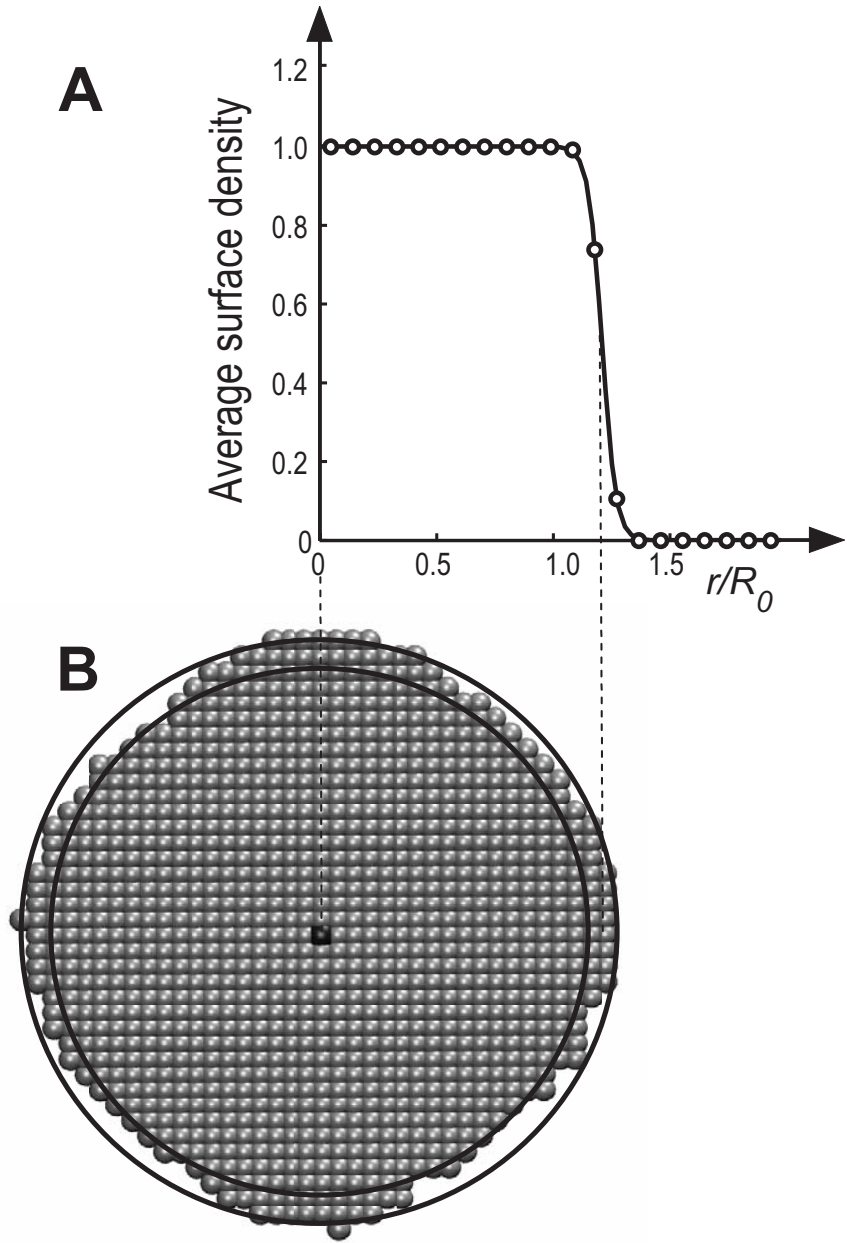

**Figure 5** (A). Definition of $r_0$, the radius of the interfacial contact. The contact surface between the fusing aggregates is approximated by a circle. The surface density of model cells in this surface is plotted vs. the distance from the center of the circle. The density is fitted by the function given in Eq. 11. The disk radius is defined as the radial coordinate at which this function drops to half of its maximum. (B). The contact disk obtained in $7\times10^5$ MCS.

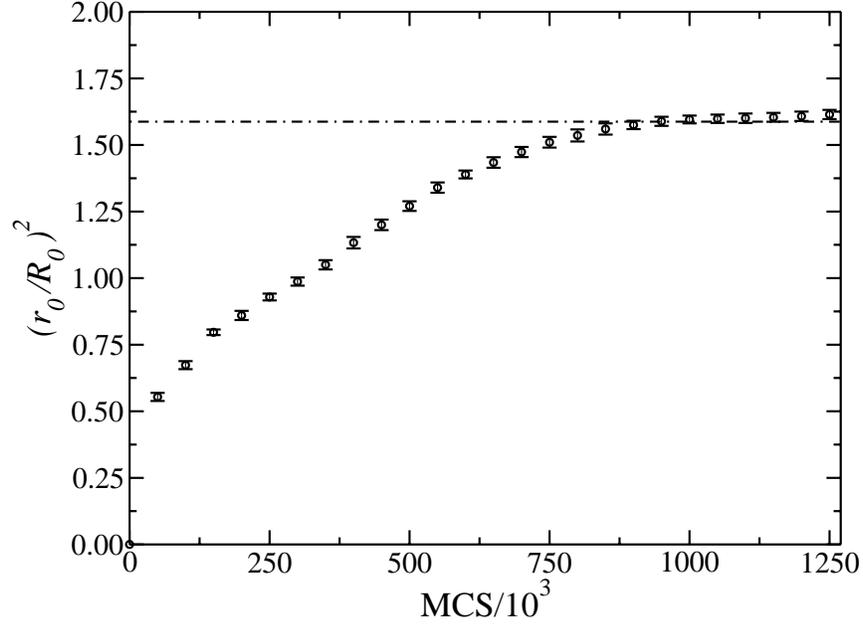

**Figure 6** Simulated evolution of $(r_0/R_0)^2$. The open circles represent mean values obtained from 30 MC simulations with interfacial tension parameter $\gamma_{cm} = 0.5$ (the simulations differed only by the seed of the random number generator). The dash-dot line shows the target value derived from volume conservation. Error bars are based on standard deviation around the mean.

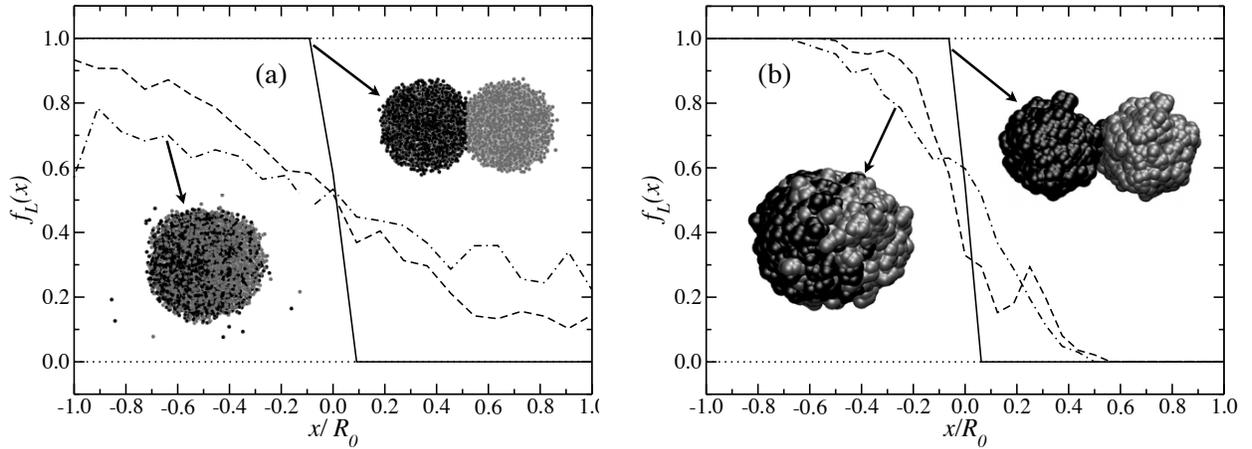

**Figure 7** The fraction $f_L(x)$ in the two CPD simulations that originate from the left aggregate for $t = 0$ (solid line), $t = 0.5t_R$ (dashed line), and $t = t_R$ (dashed dotted line) in case of simulations (a) CPD-1 and (b) CPD-10. Snapshots in both panels are shown in the initial state at $t = 0$ (upper right) and $t = t_R$ (lower left).

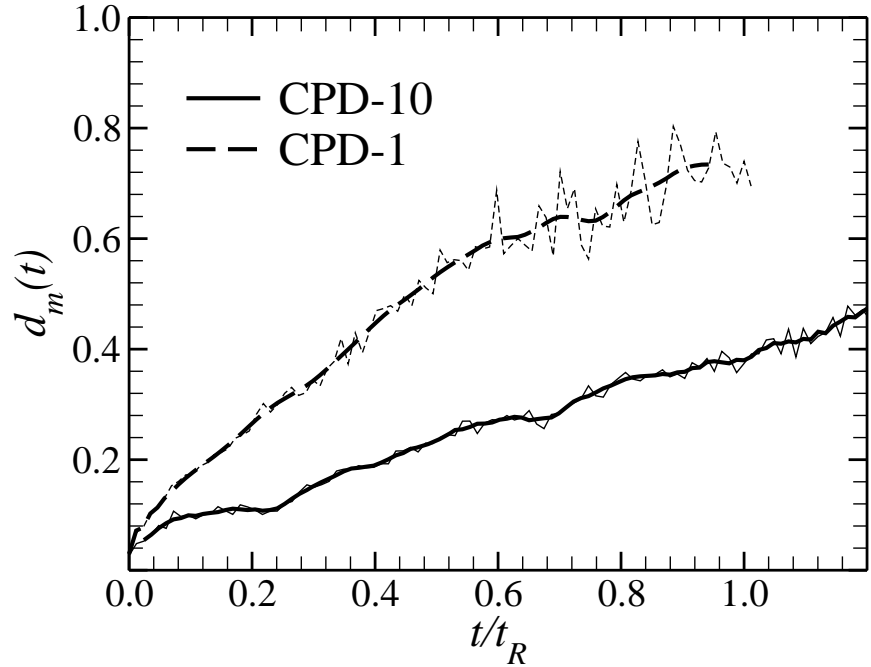

**Figure 8** Mixing parameter $d_m$ for the CPD-10 (solid line) and CPD-1 (dashed line) simulations. The fluctuations (noise) in the simulation data (thin curves) were reduced by a moving-window averaging procedure, leading to the smooth thick curves.

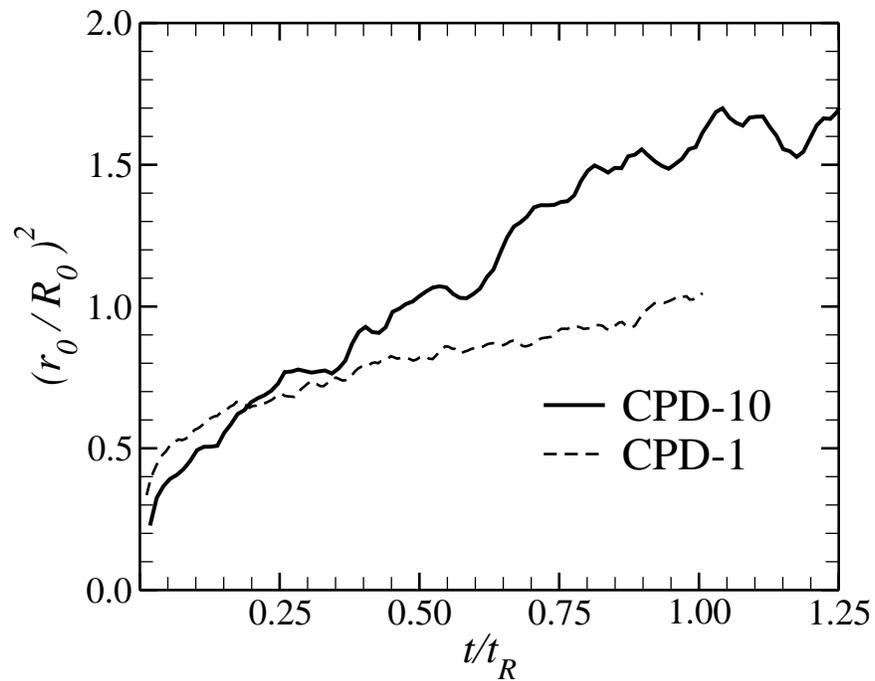

**Figure 9** Time evolution of the contact area between the two fusing cell aggregates obtained in simulations CPD-10 (solid line) and CPD-1 (dashed line).

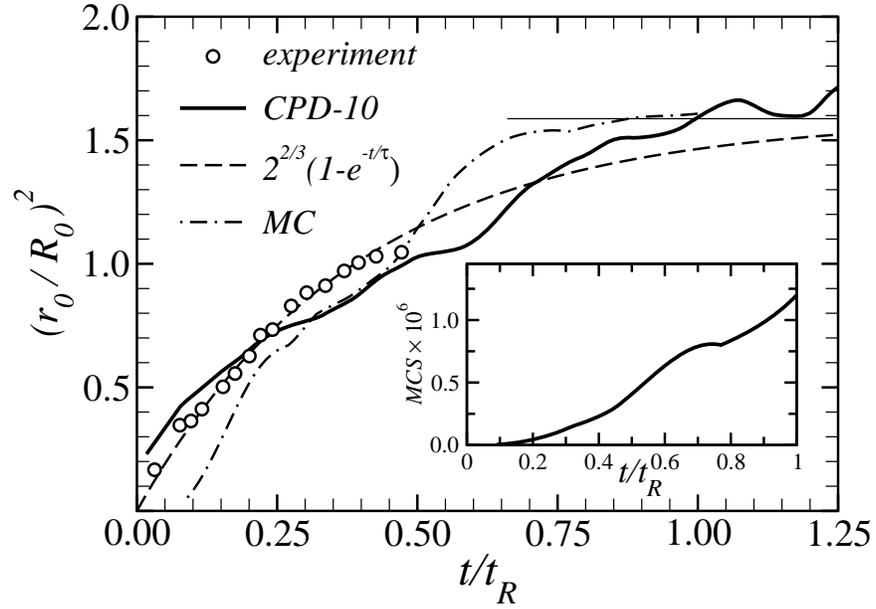

**Figure 10** Comparison of the time evolutions of the interfacial area between two fusing aggregates obtained from experiment (open circles), CPD-10 simulation (solid line), MC simulation (dot-dashed line) and theoretical prediction (dashed line). The horizontal line indicates the limiting value $2^{2/3}$ of the contact area. Inset: Correlation between the MCS and $t/t_R$ determined from shape correspondence in Fig.1.